# Simulations of β-decay of $^6$He in an Electrostatic Ion Trap

S. Vaintraub[1,2], K. Blaum[3], M. Hass[1], O. Heber[1], O. Aviv[1], M. Rappaport[1], A. Dhal[1], I. Mardor[2], A. Wolf[3]

(1) Weizmann Institute of Science - Rehovot, Israel
(2) Soreq NRC - Yavne, Israel
(3) Max-Planck-Institut für Kernphysik - Heidelberg, Germany

E-mail: sergey.vaintraub@weizmann.ac.il

**Abstract**
Trapped radioactive atoms present exciting opportunities for the study of fundamental interactions and symmetries. For example, detecting beta decay in a trap can probe the minute experimental signal that originates from possible tensor or scalar terms in the weak interaction. Such scalar or tensor terms affect, e.g., the angular correlation between a neutrino and an electron in the beta-decay process, thus probing new physics of "beyond-the-standard-model" nature.
The present system focuses on a novel use of an innovative ion trapping device, the Electrostatic Ion Beam Trap (EIBT). Such a trap has not been previously considered for Fundamental Interaction studies and exhibits potentially very significant advantages over other schemes. These advantages include improved injection efficiency of the radionuclide under study, an extended field-free region, ion-beam kinematics for better efficiency and ease-of-operation and the potential for a much larger solid angle for the electron and recoiling atom counters. . The β-decay of trapped $^6$He is discussed and preliminary Monte-Carlo (MC) simulation and error-analysis considerations are presented.

## 1. Theory

The experimental setup and description of the EIBT trap are discussed elsewhere [1]. In this short paper we will concentrate on MC-simulation of β-decay inside the trap. In the following we briefly present the theory of beta decay formalism, with special emphasis on the $^6$He β-decay. The subsequent section includes a brief MC-simulation method description and preliminary results.

The β-decay transition rate $W$ (inverse lifetime) in case of non-oriented nucleus is given by [2]

$$dW \propto \xi\left(1 + a_{\beta\nu}\frac{\vec{p}_e \cdot \vec{p}_\nu}{E_e E_\nu} + b\frac{m_e}{E_e} + ...\right) \propto \xi\left(1 + \frac{p_e}{E_e} a_{\beta\nu} \cdot \cos\theta_{e\nu} + ...\right)$$

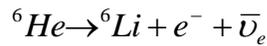
Beta-neutrino correlation coefficient

$a_{\beta\nu}$, $b$ and others are the beta decay coefficients (more coefficients also exists in case of polarized nuclei[2]). In our recent study we concentrate on the beta-neutrino correlation coefficient - $a_{\beta\nu}$, which can be measured completely unbiased in case of $^6$He beta decay, as will be shown below. The following analysis can be attuned in a similar manner to other beta decay coefficients and radionuclei.

The $^6$He nuclide has a half-life of 807 ms and is decaying into $^6$Li, electron and electron anti-neutrino

$$^6He \rightarrow {}^6Li + e^- + \bar{\upsilon}_e$$

The most general way to write the beta decay Hamiltonian [2] is:

$$H_\beta^{^6He} = \sum_{i=S,P,V,A,T} \overline{(^6Li)}\hat{O}_i(^6He)[\bar{e}\hat{O}_i(C_i + C_i'\gamma_5)\upsilon_e] + h.c.$$

where $S, P, V, A$ and $T$ stand for Scalar, Pseudo-scalar, Vector, Axial-vector and Tensor respectively. Then, the beta-neutrino correlation coefficient then can be expressed by the Gamow-Teller (GT) and Fermi (F) matrix elements along with eight coupling constants C, C' [2]:

$$\xi = |F|^2\left[|C_V|^2 + |C_S|^2 + |C_V'|^2 + |C_S'|^2\right] + |GT|^2\left[|C_T|^2 + |C_A|^2 + |C_T'|^2 + |C_A'|^2\right]$$

and

$$a\xi = |F|^2 \left[ |C_V|^2 - |C_S|^2 + |C'_V|^2 - |C'_S|^2 \right] + \frac{|GT|^2}{3} \left[ |C_T|^2 - |C_A|^2 + |C'_T|^2 - |C'_A|^2 \right]$$

The $^6$He beta decay is a $J^\pi=0^+$ to $J^\pi =1^+$ pure Gamow-Teller decay. In that case the beta-neutrino correlation coefficient can be expressed only by Axial and Tensor coupling constants in the Hamiltonian, since the Fermi matrix element is zero. In the Standard Model (SM) only the V-A interactions can occur, therefore only the axial coupling constants remain [2]:

$$|C_T|^2 = |C'_T|^2 = 0 \qquad |C_A|^2 = |C'_A|^2 = 1$$

Consequently the transition rate for pure Gamow-Teller decays in the SM frame has the following simple form,

$$dW_{SM}(\theta_{ev}) \propto \left( 1 - \frac{p_e}{3E_e} \cos\theta_{ev} \right)$$

where the squared GT matrix element can be brought out of brackets, hence its absolute value is not important. On the contrary, in a general case, beta decay may be neither a pure Gamow-Teller nor a pure Fermi decay. As a result any beta decay coefficient in the transition rate equation would be matrix elements depended, thus nuclear structure theory depended. However, even in case of pure GT-decay of $^6$He the second order corrections become important when precision measurements are below 1% [3, 4] and should be taken into account in the simulations.

## 2. Monte-Carlo Simulations

The simulations of the beta decay provide the energy and angular dependences of outgoing particles by considerations of the total 3-body energy and momentum conservation. The second order corrections introduce additional energy and angular dependences for the beta decay parameters. However for an energy average, in the $^6$He beta-neutrino correlation coefficient $a_{\beta\nu}$ those dependences are rather weak [4]. Therefore the beta decay rate for $^6$He has linear dependence to the beta-neutrino correlation coefficient $a_{\beta\nu}$. Thus *any distribution* will be linear with regards to $a_{\beta\nu}$. The extraction analysis of the beta-neutrino correlation coefficient is based on the comparison between the experimental results and those obtained for two sets of realistic Monte Carlo (MC) simulations considering pure axial ($a_{\beta\nu} =-1/3$) and pure tensor ($a_{\beta\nu} = +1/3$) couplings – here labeled as the "basis function" and produced with high statistics. $a_{\beta\nu}$ is then deduced from a fit of the experimental spectrum to a linear combination of the two basis functions as defined above. To mimic the eventual experimental data, a relatively low statistics simulation with $a_{\beta\nu}=-1/3$ was produced and fitted by a least squared method. The results are shown in Fig. 1, together with the two "basis" simulations (Figure 1).

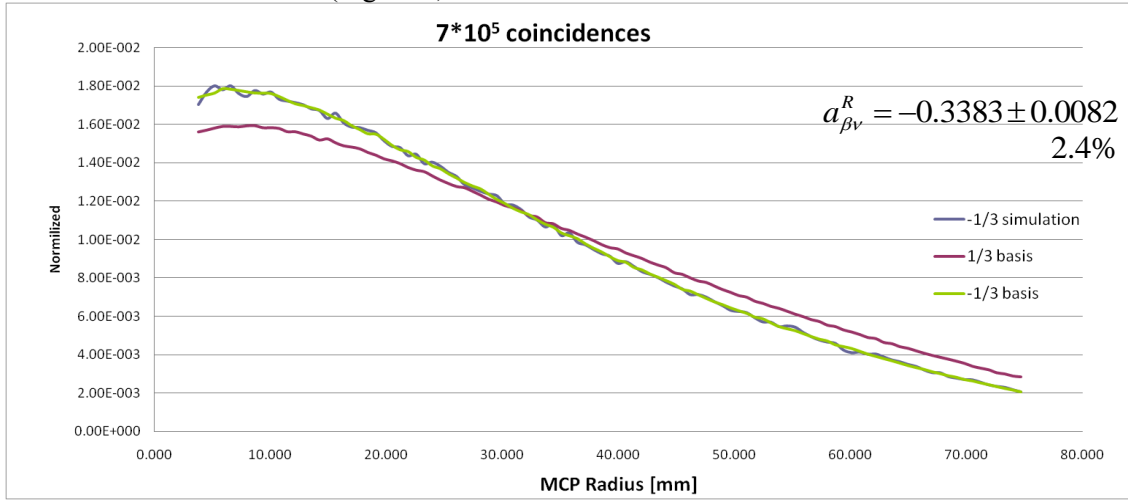

$$a^R_{\beta\nu} = -0.3383 \pm 0.0082$$
2.4%

**Figure 1** *Extraction of the β–ν correlation coefficient from the $^6$Li recoil distribution on the MCP.*

The same approach as for 1D distribution may be applied to 2D distribution (see Figure 2). In the present example the Time of Flight (TOF) of $^6$Li recoils is plotted against its distribution on Multi-Channel-Plate (MCP radius). The 2D distribution fitting significantly increases the accuracy and precision of the extraction analysis.

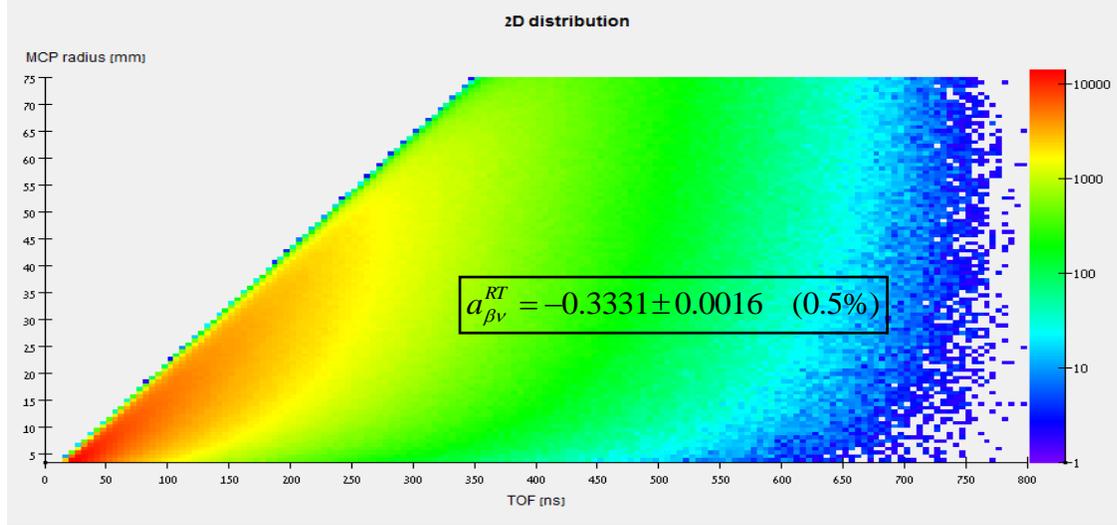

**Figure 2.** *A 2-d correlation plot of the time-of-flight vs. position on the MCP of the recoiling $^6$Li ions*

In total there are six different measurable distributions: TOF and (X,Y) positions of the lithium ions and the energy and (X',Y') positions of electrons. In principle, the complete set of measurements can be used in the extraction analysis of $a_{\beta\nu}$. However these distributions are highly correlated as, shown in Figure 2. One proper way to add all this information together is by fitting 2D distributions of any two distributions (MCP radius = R and TOF = T, for example) and each of them separately and find the covariance between the two. Then by filling the covariance matrix "V" for all pairs of distributions, one gets the average value for the correlation beta-neutrino coefficient and its average error, as shown below.

$$V = \begin{pmatrix} \sigma_R^2 & \text{cov}(\sigma_R,\sigma_T) & \dots \\ \text{cov}(\sigma_R,\sigma_T) & \sigma_T^2 & \dots \\ \dots & \dots & \dots \end{pmatrix}; \quad W = \begin{pmatrix} 1 \\ 1 \\ \dots \end{pmatrix}; \quad A = \begin{pmatrix} \alpha_R \\ \alpha_T \\ \dots \end{pmatrix} \quad \begin{matrix} \overline{\alpha} = \sigma_{\overline{\alpha}}^2 (W^T V^{-1} A) \\ \\ \sigma_{\overline{\alpha}}^2 = (W^T V^{-1} W)^{-1} \end{matrix}$$

3. **Conclusions**

The examples above present preliminary results; a full analysis will include the complete covariance matrix, taking into account additional measurable parameters of the experiment, like energy and position of the emitted electrons. This approach is instrumental in order to assess the number of collected events needed for any given precision as well as to understand in detail the possible systematic errors.